*Review*

# Setting the Stage for Habitable Planets

## Guillermo Gonzalez

Department of Physics and Astronomy, Ball State University, Muncie, IN 47306, USA;
E-Mail: ggonzalez@bsu.edu; Tel.: +1-765-285-4719; Fax: +1-765-285-5674

**Abstract:** Our understanding of the processes that are relevant to the formation and maintenance of habitable planetary systems is advancing at a rapid pace, both from observation and theory. The present review focuses on recent research that bears on this topic and includes discussions of processes occurring in astrophysical, geophysical and climatic contexts, as well as the temporal evolution of planetary habitability. Special attention is given to recent observations of exoplanets and their host stars and the theories proposed to explain the observed trends. Recent theories about the early evolution of the Solar System and how they relate to its habitability are also summarized. Unresolved issues requiring additional research are pointed out, and a framework is provided for estimating the number of habitable planets in the Universe.

## 1. Introduction

The circumstellar habitable zone (CHZ) has served as a unifying concept in astrobiology for several decades, but the broader astrophysical context of habitability (e.g., the origin and distribution of the elements that go into forming planets) requires that we also consider galactic-scale habitability (galactic habitable zone (GHZ)) and cosmic-scale habitability (cosmic habitable age (CHA)) [1]. However, these divisions of habitability are somewhat arbitrary, and some processes relevant to habitability cannot be easily confined to just one type of zone. In particular, the steps leading to the formation of a habitable planet encompass all of cosmic history when the astrophysical sources of the elements that go into making it are considered. While we make extensive use of these habitable zone divisions in the present review, we will also point out their inadequacies.

It is important to clearly define what we mean by the term "habitable planet". Some will object that the word "planet" is too restrictive. In principle, a habitable environment might be located on a non-planetary body. Perhaps the term "habitable body" would be preferred, which could include such possibly habitable objects as dwarf planets and moons. However, the possibility of a habitable environment apart from the surface or near the surface of a rocky (terrestrial) planet remains highly speculative, and we already know that a terrestrial planet can be habitable (e.g., Earth). However, we do include discussion in the present review of the possibility of habitable moons, since such bodies are still "planet-like", even though they are not technically planets.

Second, the word "habitable" can take on a number of meanings. As employed in the present review, habitability refers to the capability of starting life and sustaining it. A habitable planet could include an environment capable of supporting only one or two extremophile species in low abundance or a lush and diverse biosphere. The word "life" can also take on a number of meanings. Sometimes, life is divided into "simple" and "complex" or "single-celled" and "animal", with corresponding estimates of the CHZ boundaries for each type of life that can originate and be sustained [2]. Even alternative kinds of chemistry upon which life might be based are sometimes considered, but such explorations are highly speculative and have not produced viable instances. In the present review, we will therefore restrict the discussions to carbon- and water-based life residing on the surface or near-surface of a terrestrial-type planet. We focus on the habitability requirements for complex metazoan life (given its greater sensitivity to astrophysical, geophysical and climatological factors).

The conditions required for the continued existence of life on a planet are probably different and less specific than the conditions required for its origin. For example, some origin-of-life scenarios require dry land to be present and go through dry-wet cycles, something not possible on a "water world" [3]. A planet that fails to be in the required "window" for the origin of life early in its history could later be within the CHZ, but lack life [4]. Contrariwise, a planet that was once habitable can go through a sterilization phase; it could return to a habitable state, but lack life. These cases serve to remind us that a planet can be habitable, while at the same time being uninhabited. Therefore, to determine if a planet is habitable at some specific time, it is necessary to follow its detailed evolution from the time of its formation.

Following [5,6], it is helpful to define four classes of potentially habitable planets. Class I habitats maintain liquid water on the surface and are exposed to light form the host star. Earth is an example of a Class I habitat. A Class II habitat begins with surface water, but loses it within a few billion years. Mars and Venus are in Class II. Class III habitats have a subsurface water ocean that interacts with the silicate interior. Europa is an example of this class. Class IV habitats have liquid water above an ice layer, separating the liquid water from the silicate core. Ganymede and Callisto are probably examples of a Class IV habitat. In the present review, we will only consider Class I habitats, as this is the most likely class to be potentially habitable.

The field of astrobiology has developed at an ever-increasing pace in the last decade, and it continues to attract researchers with diverse backgrounds. This is needed, as astrobiology covers highly diverse topics, ranging from biology to cosmology. It is very difficult to present a complete review of astrobiology. Even within the "narrow" focus of the present short review on the conditions required for planetary habitability, it is not possible to cover all the relevant research. Given this, I will limit the review to those recent studies that are most likely to have the greatest influence on future research in this area.

The present review is organized as follows. We begin by discussing the CHZ, GHZ and CHA concepts as they relate to setting the preconditions for the formation of habitable planets. This is followed by reviews of the latest observational and theoretical research on exoplanets and the early history of the Solar System. We close with a suggested framework for estimating the number of habitable planets in the Universe.

## 2. The CHZ, GHZ and CHA Concepts and Habitable Planet Formation

### 2.1. The Circumstellar Habitable Zone (CHZ)

In the most basic version, the CHZ is defined assuming an energy balance between a terrestrial planet and its host star. A terrestrial-type planet within the CHZ of its host star is considered potentially habitable if it can maintain liquid water on its surface. More often, the circumstellar continuously habitable zone (CCHZ) is considered, which is the region of continuous habitability over at least a few billions of years. Recent studies of the CCHZ and CHZ build on the seminal work of James Kasting and collaborators of 20 years ago [7]. They calculated a set of CHZ models for the Sun and stars of other stellar spectral types. They differed from earlier models primarily by including the carbonate silicate feedback cycle and a more accurate treatment of energy transport in the atmosphere. Kasting *et al.* [7] defined the inner boundary of the CHZ in multiple ways. One is based on the "moist greenhouse". In this process, water gets into the stratosphere, where it is dissociated by solar UV radiation, and the H atoms are lost from the top of the atmosphere. A second definition for the inner boundary is based on the runaway greenhouse effect. They calculated the outer boundary according to the maximum possible $CO_2$ greenhouse or the increase of planetary albedo, due to the formation of $CO_2$ clouds. The inner and outer boundaries were also estimated from the observed states of Venus and Mars, respectively. Their most restrictive case has inner and outer boundaries of 0.95 and 1.37 AUs, respectively.

Many papers on the CHZ have been published since 1993. Several efforts have focused on improving the modeling of radiative processes in a planet's atmosphere. For example, [6,8] have found, contrary to prior work, that the formation of $CO_2$ clouds near the outer boundary of the CHZ produces little to no additional greenhouse warming. In addition, Forget [6] showed that the greenhouse warming contributed by gas-phase $CO_2$ was probably overestimated by Kasting *et al.* [7]. Both these corrections lead to a smaller outer radius for the CHZ than had been previously calculated. Shields *et al.* [9] find that ice and snow albedo feedback is dependent upon the host star's spectrum and leads to a greater climate sensitivity to changes in the host star for hotter stars; this results in an increase in the inner radius of the CHZ around F and G stars. Kopparapu *et al.* [10,11] employed cloud-free 1D climate models for F to M spectral types and found that the inner edge radius is increased relative to previous calculations; for the Solar System, they find that the moist greenhouse inner edge is only 0.01 AU inside the Earth's orbit! Not surprisingly, a continuing source of uncertainty in the calculations of CHZ boundaries (as well as climate sensitivity research for Earth) concerns the treatment of clouds.

The star in a planetary system supplies the energy needed to maintain liquid water on the surface of a terrestrial planet in the CHZ, and this is the only role it plays in most studies of the CHZ. However, it has other important effects on the terrestrial planets that are likely important for habitability. These include its gravitational influences, UV radiation, particle radiation and the stellar wind and irradiance variations.

Another constraint on the boundaries of the CHZ comes about when the positive and negative effects of the UV radiation from the host star are included [12]. The inner boundary of the "UV-CHZ" is set by the maximum UV flux that DNA can tolerate, and the outer boundary is set by the minimum required UV flux for biochemical reactions. Guo *et al.* [13] find that only stars between about 0.6 and 1.9 $M_s$ will be within both the traditional CHZ and the UV-CHZ.

Particle radiation escaping from a star is produced in its chromosphere and corona and can greatly increase during coronal mass ejection (CME) events, which are often associated with flares. Ionizing radiation (particles and short wavelength electromagnetic radiation) can damage the ozone UV shield on a planet with an oxygen-rich atmosphere [14]. In the Solar System, the solar wind modulates the cosmic ray flux (from beyond the Solar System) and affects their flux at the Earth [15]. There is some evidence for a link between cosmic ray flux and climate, but it remains controversial [16]. Recently, a spike in the atmospheric C-14 level was found to have occurred in AD 774–775; it has been attributed to a very strong solar proton event (more intense than any similar event measured during the satellite era or from paleo records going back several thousand years) and should have had moderate ozone depletion [17]. It corresponded to a fluence of protons with energy >30 MeV near $5 \times 10^{10}$ cm$^{-2}$ [18]. Stronger solar flares and related phenomena of increasing magnitude occur with decreasing frequency, but radionuclide evidence from lunar rocks indicates an upper limit on their magnitude up to a timescale of a Myr (megayear) [19].

Very large transient increases in the radiation from Sun-like stars, termed *superflares*, have been suspected for nearly 20 years, but high quality statistics have only become available in the last few years from observations made with Kepler. The distribution of the superflare occurrence rate with energy follows a power law form similar to that observed for solar flares, and the occurrence rate for flares with energy in the rate $10^{34}$ to $10^{35}$ erg ($10^7$ erg = 1 Joule) is estimated to be once in 800 to 5000 years, and the strongest flares observed are about a million times as energetic as the strongest observed solar flare [20]; the most energetic solar flare to occur during the satellite era produced about $10^{32}$ erg [21]. Stars that rotate more slowly produce superflares less frequently than fast rotators, but they are comparable in energy [22]. It had already been known for some time that the gradual slowing of the rotation of a sun-like star as it ages corresponds with the decline of its activity. A recent surprising discovery is the occurrence of superflares in some early F and A stars (with typical energies of $10^{35}$ and $10^{36}$ erg, respectively) [23].

Compared to the quiescent luminosity of a star, flares in K and M dwarfs are relatively more energetic than flares in G dwarfs. This is relevant to habitability, because the location of the traditional CHZ is determined by the quiescent luminosity of a star. Therefore, a planet in the CHZ of an M dwarf will be subjected to a much greater flux of intermittent ionizing radiation than a planet in the CHZ of a G dwarf. Both the particle and electromagnetic ionizing radiation can damage the ozone layer of an Earth-like atmosphere, and a portion of the energy from the X-ray and gamma ray photons can be redistributed to biologically damaging UV radiation that reaches the surface of a planet [24]. In addition,

Sun-like stars that have more frequent flares also have larger irradiance variations on multiple timescales [25], likely causing larger climate variations.

Ionizing radiation and stellar winds can remove the atmosphere from an Earth-like planet in the CHZ of a low mass star, but the timescale for its loss depends on several factors, including the masses of the planet and its atmosphere, its distance from the host star and the strength of its magnetic field [26]. The strength of a terrestrial planet's magnetic field, in turn, depends on several factors, including the presence of liquid iron in its core and its rotation period [27]. The rotation of a planet in the CHZ of an M dwarf will be slowed due to the action of the tides from its host star, leading to a weakening of its magnetic field on relatively short timescales and, hence, more rapid loss of its water and part of its atmosphere [27,28].

A planet orbiting a low mass star can become tidally locked on relatively short timescales [29]. As a planet approaches rotational synchronization, it also undergoes "tilt erosion", which results in a very low obliquity for the planet's rotation axis [29]. This effectively eliminates seasonal variations on the planet and makes it more likely that its water will be locked up on its night side. This poses a number of potential problems for the habitability of the planet. For example, if the temperature at any location on the planet (such as its night side and/or at high elevation) is permanently below the freezing point of water, then the water becomes trapped there (an "ice trap"), resulting in a dry world [30]. Simulations of tidally-locked terrestrial-type planets have shown that a thick $CO_2$ atmosphere can avoid freezing temperatures on the night side [31]; of course, such a planet would not support animal life, which requires a low-$CO_2$, high-$O_2$ atmosphere. Moreover, planet formation models indicate that terrestrial planets formed in the CHZ of M dwarf stars might be deficient in volatiles [32]. Earth avoids the potential catastrophe of cold traps, due to its short rotation period, relatively thick atmosphere and oceans and modest obliquity angle; if Earth's obliquity angle were close to zero degrees, for example, it would be in danger of having cold traps at its poles. Excessive tidal heating, like the case of Io in our Solar System, can cause a planet approaching rotational synchronization to lose most of its hydrogen and, thus, water; such planets have been dubbed "Tidal Venuses" [33]. Tidal locking can also create new climate instabilities not previously considered [34].

More researchers are beginning to tackle the question of exomoon habitability, that is, the habitability of a moon that orbits a Jovian planet within the CHZ of its host star [35–37]. Such a world faces a number of severe challenges to its habitability. First, it is likely that Ganymede is near the maximum mass for a moon that forms around a Jovian planet like Jupiter. Models of *in situ* formation of moons show, however, that more massive planets are accompanied by more massive moons, and that Mars-size to Earth-size moons could form around Jovian planets near the upper end of their mass range (~13 Jupiter masses) [35]. Such massive Jovian planets are uncommon. Second, a moon will undergo rotational synchronization relatively quickly, resulting in slower rotation compared to Earth; it would likely have a weaker magnetic field. It could be protected by the host planet's magnetic field, but it would then experience high particle radiation levels [37]. Third, a moon orbiting a Jovian planet in the CHZ would likely have been brought there by its migrating host, having formed farther from the host star, beyond the "frost-line"; its composition would likely reflect that of the large icy moons in the outer Solar System. If the Jovian planet migration is due to torque from the protoplanetary disk, any accompanying moons are likely to survive its trek to the CHZ [38]. However, if a planet's journey to the CHZ is the result of planet-planet scattering, which is more likely for the more massive Jovian planets, then its moons are unlikely to survive [39]. In both cases, the probability of a moon remaining with its Jovian planet host is smaller if the moon is farther from the planet. Fourth, for certain combinations of planet-moon parameters, the tidal heating a moon experiences can be severe, as Io in our Solar System well illustrates. Fifth, the gravitational focusing effect of the Jovian planet host will make impacts on its moons both more frequent and more energetic than they otherwise would be.

A largely unexplored effect on the location of the outer boundary of the CHZ is the radial dependence of the asteroid and comet impact rates on a terrestrial planet. The asteroid impact rate on Mars from meteorites impacting with energies greater than a megaton is estimated to be about five times that on Earth, determined from crater counts [40]. This should not be surprising, given Mars' proximity to the asteroid belt. Indeed, [41] have performed simulations of asteroid and comet impacts on the terrestrial planets, finding that Mars receives many more impacts than Earth, despite its smaller size; they also find that fewer comets impact Mars, but the numbers are comparable when the planets' different sizes are taken into account. Of course, the asteroid impact threat on planets in the CHZs of exoplanetary systems will depend on the properties of the asteroid belts in those systems; Martin and Livio [42] argue that the formation of an asteroid belt in a planetary system is most sensitive to the location of the snow line and whether giant planets in the system undergo a large amount of migration. The details of the formation of the asteroid belt in the Solar System are reviewed in Section 3.3 below.

The above discussion of the CHZ is relevant to the case of planets orbiting a single star. The observed fraction of stars in binary and higher order multiple star systems in the solar neighborhood is about 46% [43], which is smaller than previous estimates, such as those of Duquennoy and Mayor [44]. Ragvadan *et al.* [43] also show that metal-poor stars are more likely to be accompanied by stellar companions than solar-metallicity stars. The situation for planetary habitability changes considerably when binary and multiple stars are considered. Both dynamical and radiation fluxes need to be taken into account. The two types of planetary orbits usually considered in binary systems are P-type (the planet orbits both binary components) and S-type (the planet orbits one binary component). Several studies have explored planetary habitability in both types of binaries primarily through numerical means [45–49] and also analytically [50].

## 2.2. The Galactic Habitable Zone (GHZ)

Gonzalez *et al.* [51] and Lineweaver [52] introduced the galactic habitable zone (GHZ) concept. The GHZ describes the regions of the Milky Way most likely to contain habitable planetary systems. Two classes of processes set its boundaries: the formation of Earth-like planets and threats to life. Another possible way to describe these processes is the following: setting the initial conditions for the formation of a habitable planetary system and the ability of a planetary system to sustain (complex) life over several billions of years. Gonzalez *et al.* examined the first class within the context of galactic chemical evolution. The second class includes gamma ray bursts, supernovae, comet showers and encounters with interstellar clouds [1,53,54]. Lineweaver [55] provided a more quantitative treatment of the GHZ from numerical galactic chemical evolution models, which they used to study the effects of metallicity on the formation of Earth-like planets and the distribution of supernovae.

Gonzalez *et al.* [51] assumed that the typical mass of a terrestrial planet scales with metallicity raised to the 1.5 power. Lineweaver [52] assumed, instead, that the probability of forming an Earth-like planet is linearly proportional to metallicity and drops to zero for a metallicity value of 1/10 solar. Lineweaver [52] also assumed that the probability of destroying Earths from the destabilizing effects of migrating giant planets increases linearly with the incidence of hot Jupiters. All these assumptions need to be revisited in the light of recent exoplanet observations and theoretical work. Empirical constraints on the dependencies of terrestrial and giant planet incidences on host star metallicity will be discussed in Section 3.1, and simulations of planetary system formation and evolution will be discussed in Section 3.2.

The dependence of planet formation on metallicity translates into a variation of planet formation with location and time in the Milky Way Galaxy (and the broader Universe). For example, the Milky Way, like other spiral galaxies, exhibits a radial disk metallicity gradient in the sense that the outer disk is more metal-poor than the inner disk. Thus, planets should form more easily in the inner galaxy. While this is an extrapolation based partly on local observations, there is some empirical support for this claim [56]. In addition, the metallicity of the disk gas has been increasing since the Milky Way formed. This implies that the rate of planet formation should be increasing with time. These trends help to define the present day GHZ and its evolution.

Metallicity varies with time and location in the Milky Way Galaxy. The galaxy is often subdivided into the halo, bulge and disk, each component characterized by the distribution, dynamics and nature of its matter content. While occupying the largest volume, the halo contains only old metal-poor stars. The bulge has the highest density of stars, which range in metallicity from about one-thirtieth solar to about three times solar [57]. Most of the bulge stars are within a few kiloparsecs (kpc) of the galactic center and have large orbital inclinations relative to the disk.

As a result of the continuing star formation in the thin disk, the metallicity of the gas, and thus of the stars that form from it, has steadily increased. The observational signature of the metallicity evolution is termed the age-metallicity relation (AMR). The rate of star formation is not spatially uniform in the disk. Star formation has preceded more quickly in the inner disk of the galaxy, the observational evidence of which is the radial metallicity gradient, $\alpha$.

Within the context of defining the boundaries of the GHZ, $\alpha$ is the most important spatial metallicity trend in the galaxy. Many estimates of $\alpha$ have been published over the past two decades using a variety of chemical abundance tracers. The disk metallicity tracers can be divided into two groups: "zero-age" and old objects. The zero-age objects include H II regions, B stars, Classical Cepheids ("Cepheids" from here on), young dwarf stars and young open clusters. As the name implies, zero-age objects were formed very recently, and hence, their compositions should be representative of their local environment. They are no more than a few hundred Myr old. Old tracers include K giants, old dwarf stars, old open clusters and planetary nebulae. These objects are a few Gyr (gigayears) to 12 Gyr in age. The advantage of employing zero-age objects to determine the value of $\alpha$ is that they have not had much time to wander far in radial position since their birth. We can denote the present value of $\alpha$ as $\alpha_0$.

The reliability of chemical abundance determination depends on both the type of object and on the element being studied. Chemical abundances can be determined reliably for a wide range of elements for Cepheids, K giants and Sun-like stars (whether in the field or in open clusters). In particular, the abundant metals, Si and Fe (and somewhat less precisely, O), can be measured reliably in these stars; reliable O abundances have also been determined for B stars. These types of objects also tend to be more uniform in their properties. In contrast, relatively few elements are measurable in the spectra of H II regions and planetary nebulae, but O is one of them. Moreover, the methods employed to derive nebular abundances are based on assumptions that are not as well justified as those that make use of line formation in stellar atmospheres. In particular, H II regions can have abundance, density and temperature variations that are not well modeled; the correction for condensation onto grains is not always well known, and dust reddening must be corrected for [58]. The estimate of $\alpha_0$ for O using H II regions is −0.043 dex·kpc$^{-1}$ [59].

Cepheids and open clusters have another advantage over other tracers: their distances can be measured accurately over a large fraction of the galaxy's radial extent. This is important because $\alpha_0$ is best determined when the span in the mean galactocentric radius, $R_m$, is greatest. Open clusters have the additional advantage that their ages can be accurately determined, permitting analysis of $\alpha$ for a range of ages. The average value of $\alpha$ from recent studies of open clusters is about −0.06 dex·kpc$^{-1}$ (see Lemasle *et al.* [60] for a summary of recent measurements). Empirical determinations of the time derivative of $\alpha$ remain very uncertain, but its absolute value is probably less than 0.01 dex·kpc$^{-1}$ Gyr$^{-1}$. Recent determinations of $\alpha_0$ using Cepheids observed over the range $R_m$ = 4 to 17 kpc have converged on a value near −0.07 dex·kpc$^{-1}$ [60,61].

The present metallicity of the solar neighborhood had been, until recently, a difficult number to pin down. Unless a sample is carefully prepared to contain only young objects, it can be contaminated by old objects visiting the solar neighborhood from distant regions of the galaxy. Recently, abundances of several elements have been carefully determined with high accuracy in a sample of 29 early B stars in the solar neighborhood, finding Si and Fe abundances nearly identical to those of the present Sun [62]. These results also imply that the interstellar medium (ISM) out of which the B stars formed must have been homogeneous and well-mixed. Consistent with the findings of Nieva and Przybilla [62], Nittler [63], who employed measurements of isotope ratios in presolar grains, argued that supernova ejecta do not produce more than one percent inhomogeneity in the interstellar medium.

The close agreement between the present solar abundances and the B star abundances, however, is only a coincidence. The original metallicity of the Sun was higher than the currently measured value by about 0.04 dex, due to the effects of atomic diffusion in its atmosphere. Furthermore, the metallicity of the ISM is steadily increasing, due to galactic chemical evolution (GCE). A rate of increase of [Fe/H] of 0.017 dex·Gyr$^{-1}$ has been determined from detailed spectroscopic observations of nearby F and G dwarf stars [64]. In order to account for the relatively high initial metallicity of the 4.6 Gyr old Sun compared to the present local ISM, Nieva and Przybilla [62] estimate that the Sun has migrated outward in the galaxy by about five to 6 kpc since it formed; they also cite as evidence for the solar migration the C/O ratio in the Sun compared to nearby B stars. In other words, a metallicity similar to that of the present local ISM was reached at a location well inside the present solar position in the Milky Way disk when the Sun formed.

There is little doubt that the inner regions of the Milky Way are populated by far more Jovian planets per unit volume than the solar neighborhood, both from the higher densities and the higher metallicities of the stars there. The number of terrestrial planets per unit volume should also increase towards the galactic center, but not as much as giants. Thus, stars in the inner galaxy should be relatively richer in Jovian planets compared to terrestrial planets, and the converse would be true for the outer regions. This expected trend should have significant consequences for the GHZ, since Jovian planets have multiple important effects on the habitability of a planetary system (see Sections 3.2 and 3.3).

The galactic parameters summarized above are important constraints on GCE models. A well-calibrated GCE model can be used to determine the metallicity of stars forming at any location and time in the Milky Way. It can also be used to model aspects of the Milky Way not directly observable in the distant past, such as the evolution of the supernova rate. Gowanlock *et al.* [65] present what is essentially an updated version of Lineweaver *et al.*'s [55] exploration of the GHZ for the Milky Way using more detailed GCE modeling, including Monte Carlo methods. While Gowanlock *et al.* [65] leave out several important processes likely relevant to galactic-scale habitability (e.g., Oort cloud comet perturbations [54,66], nuclear outbursts, encounters with interstellar clouds [67]), their approach is the one required to make progress in modeling the complex chemo-dynamical processes that will be adopted to define the GHZ.

The next step is the application of the GHZ concept to other nearby galaxies, including galaxies very different from the Milky Way. Carigi *et al.* [68] were the first to explore the GHZ in M31, the Andromeda Galaxy. They employed a GCE model to follow the evolution of the metallicity and the supernova rate in its disk and halo; they excluded the bulge given its high stellar density. Suthar and McKay [69] explored the GHZ for two elliptical galaxies, M32 and M87, but they only considered the effects of metallicity.

Empirical approaches to learning about galactic-scale habitability factors are also possible. In particular, many studies have sought a link between Earth's geological records and astrophysical processes occurring beyond the Solar System. Some reported to have found evidence of or influences from supernovae [70,71], the passages of the Solar System through spiral arms [72] or through interstellar clouds [67] and even the effects of gamma ray bursts [73]. However, these kinds of studies have also been subject to frequent criticisms (e.g., [74–76]). Continued research in this area is certainly warranted.

*2.3. The Cosmic Habitable Age (CHA)*

The broadest framework for discussing habitability is the CHA [1,52]. This is also the least explored habitable zone concept. It is not a spatial zone, but rather, a temporal zone of habitability over the course of the evolution of the Universe. Given that the Universe has changed so dramatically since its origin, the question naturally arises why we observe ourselves to be living during this particular time as opposed to some other time. Clearly, chemically-based life is not possible in the very early Universe before atoms formed or in the distant future, after all the stars burn out. Other considerations indicate that the boundaries of the CHA are much narrower than these extreme limits.

Progress in refining the CHA will come primarily from improvements in our understanding of the evolution of the cosmic star formation rate. The star formation rate of a galaxy depends, primarily, on the gas abundance. Star formation, in turn, determines the evolution of the supernova rate and gas phase metallicity in galaxies. However, the relationships are not straightforward, as these processes feedback on the star formation rate [77–79].

If all galaxies were just like the Milky Way, then the GHZ could just be applied to other galaxies. However, they are not; there is great variation in their properties. Galaxies differ in their Hubble types (elliptical, spiral or irregular), environment (isolated, group member or cluster member), metallicities, luminosities, masses and star formation rates. Some of these properties correlate with each other and evolve over time.

Habitable planet formation is most likely to occur over some range of metallicities. A galaxy's average metallicity increases over the history of the Universe, resulting in more probable formation of planets around each new star. The overall cosmic star formation rate has been decreasing over the history of the Universe, though local effects, such as collisions between galaxies, can temporarily enhance their star formation rates. Since an important class of astrophysical threat to life on a planet depends on the star formation rate, it is likely that the Universe overall is becoming safer.

Large surveys of galaxies over a range of redshift (e.g., [80,81]) are resulting in advances in our understanding of the relationships among these galactic properties. For example, the mass-metallicity relation, which describes the positive correlation between the total stellar mass of a galaxy and either the gas-phase or stellar metallicity, is now well-established from observations [82]; this means that low mass (and, therefore, low luminosity) galaxies are metal-poor and, hence, unlikely to contain habitable planetary systems. While there have not been any fundamental changes in understanding how they apply to the CHA since this topic was last reviewed in 2005 [1], it deserves to be revisited soon.

## 3. Learning from Exoplanets and the Solar System

### 3.1. Observed Trends among Exoplanets

Most exoplanet detections to date have come from two methods: Doppler and photometric transits. The Doppler surveys primarily target nearby field stars. The Doppler method has resulted in the detection of about 532 planets in 400 systems. The transit surveys primarily target distant stars. By far the most successful transit survey to date is NASA's space-based Kepler mission, which has detected just over 3538 exoplanet candidates as of November, 2013 [83]; sadly, its primary mission ended in 2013, due to hardware failure. Kepler has detected planets smaller in size than Earth. The similar CoRoT (COnvective ROtation and planetary Transits) mission has detected about 500 candidate transiting planets [84]. The major advantage of the nearby star surveys is that the distances to the stars are accurately known. Only when an exoplanet has been detected with both methods, its size and mass and, therefore, its density can be determined. This has been done with a few nearby stars and several dozen                                        exoplanets                                        discovered with Kepler.

Following the discoveries of the first few exoplanets (using the Doppler method), it quickly became apparent that exoplanetary systems do not resemble the Solar System; this is also true when detection biases are taken into account. Most systems either have planets in very short period orbits ("hot Jupiters") or planets in longer period eccentric orbits. For approximately the first 10 years, only massive planets ($\sim 0.5 < M_p < 12$ $M_J$) could be detected (termed "Jovian" planets, after Jupiter). The incidence of Jovian-mass planets has been found to rise steeply with increasing orbital period [85]. Refinements to the Doppler method (and also the gravitational microlensing method) have led to discoveries of many planets between the mass of Earth and Neptune ("super Earths"), as well as an abundance of Neptune mass planets. Among nearby solar-type stars, the incidence of planets of all detectable masses with periods less than 100 days is at least 50%; the planet mass histogram for periods less than 100 days peaks at a few tens of Earth masses and drops sharply for masses above 40 Earth masses [85].

The Kepler data permit the most accurate estimate of the incidence of Earth-size planets to be made. The data are most complete for orbital periods less than about 200 days; for these, the incidence of planets between one and two Earth radii and receiving between 0.25 and four times the stellar light insolation that Earth receives from the Sun is found to be 11% ± 4% of Sun-like stars [86]. Extrapolation of this estimate to orbital periods comparable to that of the Earth (around a solar analog star) has also been done [86]. However, extrapolation is always risky, and it is even more risky here. This is very likely to be an over-estimate of the incidence of habitable Earth-like planets, given that it is based on generous ranges of stellar insolation and planet size. A range of one to two Earth radii might not seem very broad, but this corresponds to a mass range of roughly one to 10 Earth masses. Figure 1 shows the dataset upon which this estimate is based. Note that only one planet within the green box in the figure is comparable in size to the Earth. Another recent analysis of the Kepler data finds that 16.5% of main sequence FGK stars have at least one planet between 0.8 and 1.25 Earth radii and periods less than 85 days [87].

**Figure 1.** Figure 4 from [86] based on Kepler observations of small planets. The green box contains the planets most similar in size and received stellar insolation to Earth.

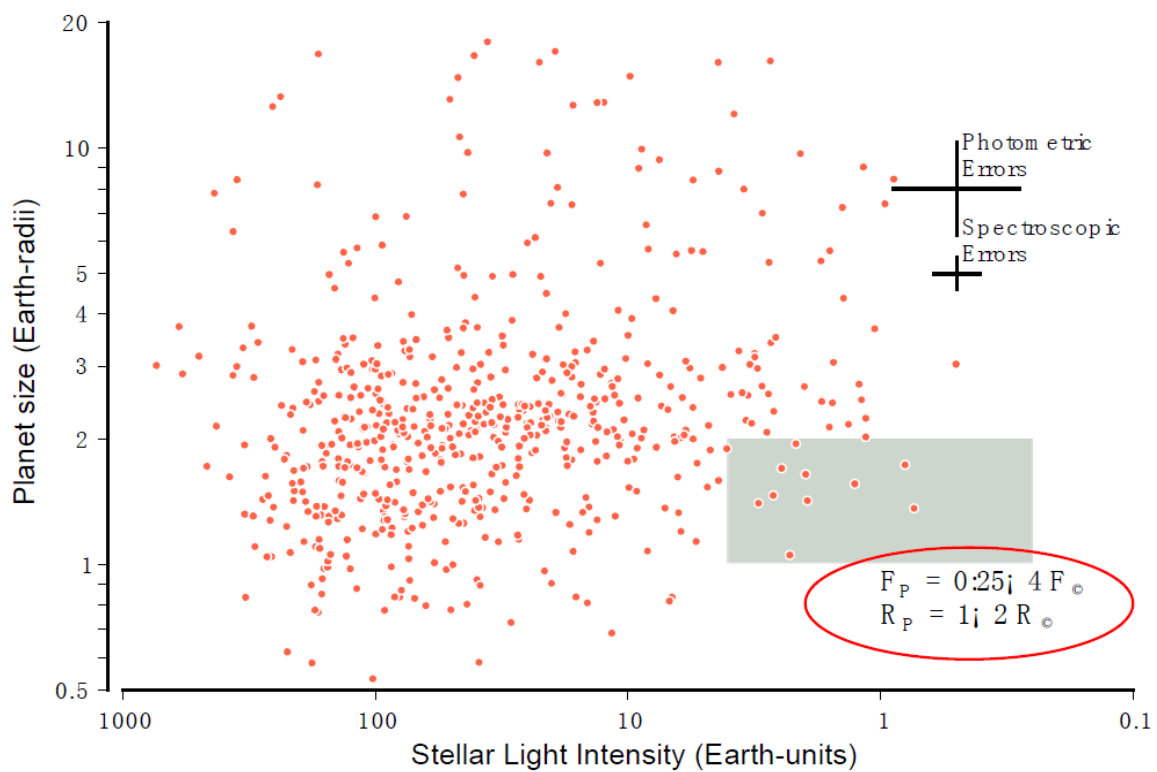

Other important trends include the mass and radius distributions of planets. The incidence of planets rises steeply with decreasing mass in the Doppler samples of nearby stars, and it rises steeply with decreasing radius in the Kepler transit survey sample; below ~2.8 Earth radii, the Kepler planet size distribution levels off [88]. When both the mass and radius are measured for an exoplanet, it is possible to say something about its composition. Figure 2 shows that the most likely compositions for the well-characterized exoplanets spans the range from metallic to gas giants, including a number of highly extended gas giants. This figure also illustrates the high sensitivity of planet mass on radius for a rocky planet composition.

The eccentricity distribution of Doppler-detected planets more massive than Saturn and with orbits beyond 1 AU peaks near 0.2, but it has a long tail extending close to 1.0 [89]. An up-to-date eccentricity distribution is shown in Figure 3; its median value is 0.18, and increasing the minimum orbital period plotted from 20 to several hundred days increases the median slightly to about 0.22. For reference, Jupiter and Saturn have orbital eccentricities of 0.048 and 0.053, respectively.

**Figure 2.** Figure 3 from [88] showing the radii and masses of well-characterized exoplanets as red open circles. Modeled mass-radius curves for various pure compositions are shown as blue curves. Solar System planets are shown as green triangles. Figure courtesy of Andrew W. Howard.

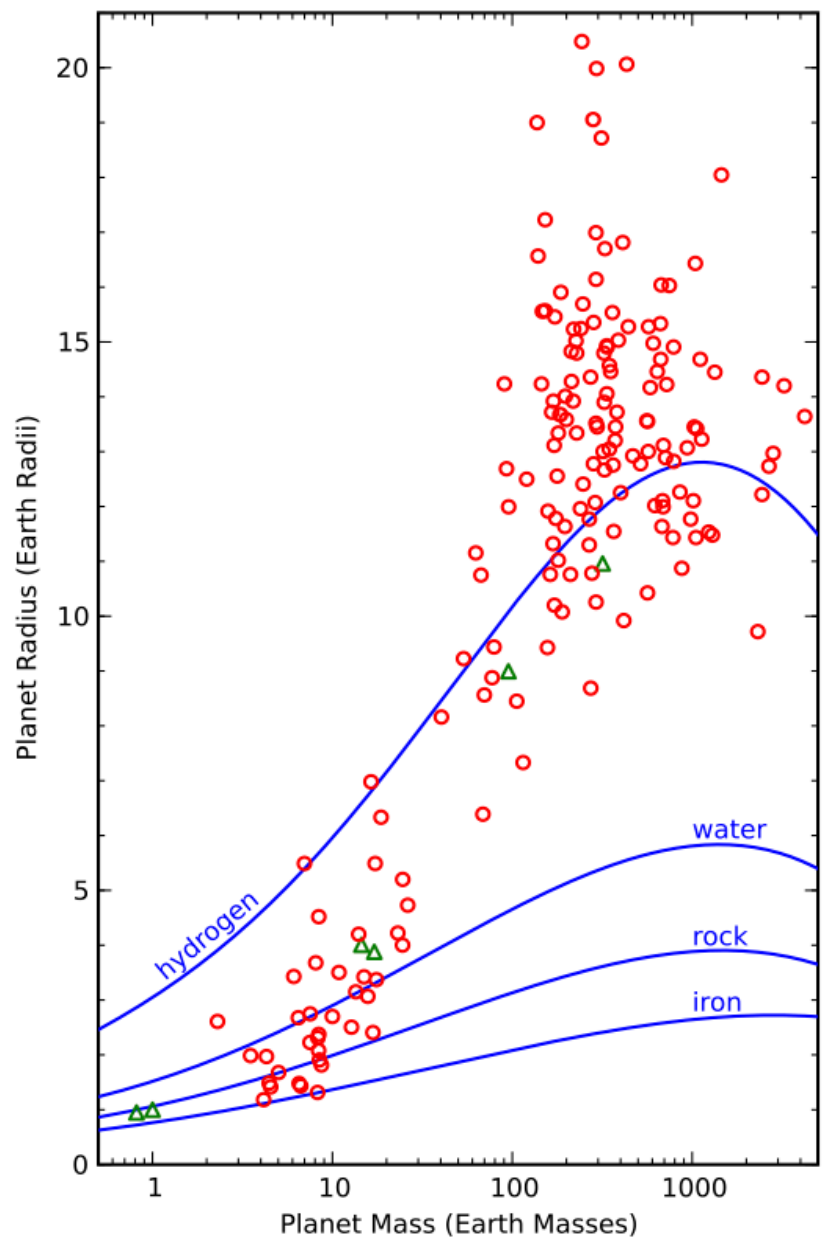

**Figure 3.** Eccentricity distribution of 394 exoplanets with orbital periods greater than 20 days from the vetted data in the Exoplanet Orbit Database. The median value of the distribution is 0.18.

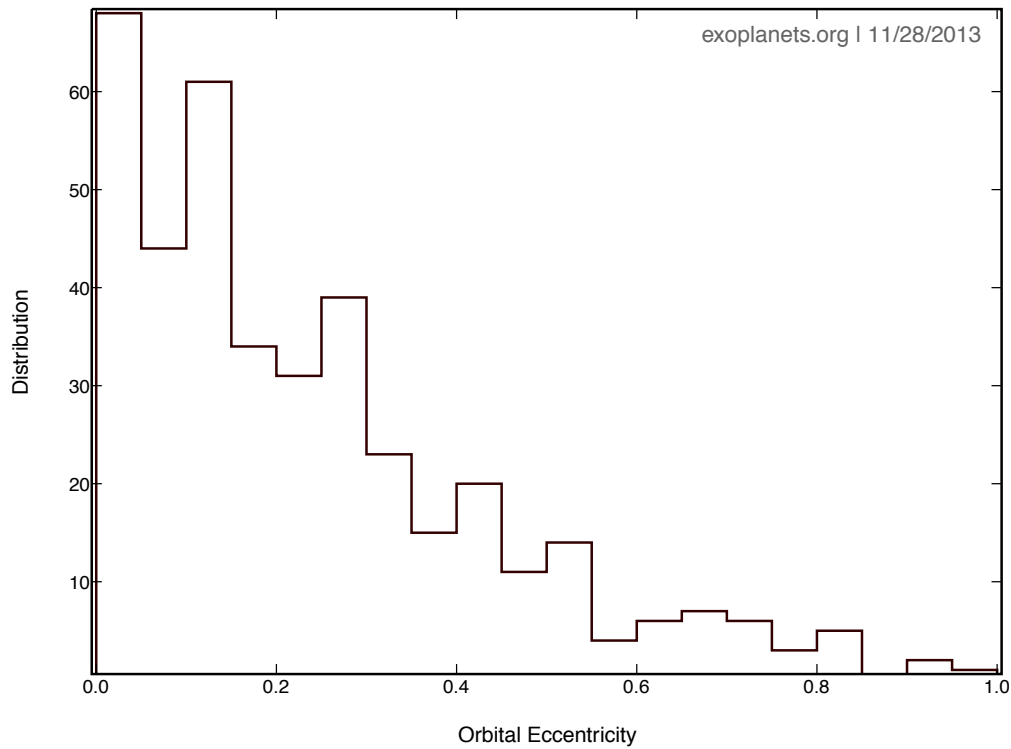

Many multiple planet systems have been discovered, both among the nearby star Doppler samples and the Kepler sample. About one-third of the ~1200 candidate planets discovered with Kepler during the first four months of observations were found to be members of multiple planet systems [90]. One of the somewhat surprising discoveries among both the nearby and Kepler samples is the existence of systems having two or more planets in or near mean motion resonances (meaning that orbital period pairs form simple ratios of small integers, such as 2:1). However, dynamicists had known that this was a theoretical possibility amongst planets prior to the discovery of the first such system (around the star, GJ 876).

Short-period planets tend to have very circular orbits, which presumably have been circularized by tidal interactions with the host stars [91]. Other star-planet dynamical interactions leading to changes in the rotation periods of the host stars have been inferred from the observed deficit of planets around fast-rotating stars in the Kepler sample [92] and smaller projected rotation velocities (vsini) among nearby stars, with Doppler-detected planets compared to stars without planets [93].

The correlation between the host star's metallicity and the presence of Doppler-detected giant planets was the first important star-planet relation to be discovered [94]. This trend has since been confirmed independently by several research groups [95,96]. The incidence of giant planets rises steeply for dwarf stars that are more metal-rich than the Sun; a power-law equation describes the relative frequency of Jovian-mass planets:

$$f_p = C\,10^{\alpha[Fe/H]} \tag{1}$$

where $C \sim 0.02$–$0.04$ and $\alpha = 1.26$–$2.94$ [97]. Assuming $\alpha = 2.94$, if a star has [Fe/H] = 0.20, then it is almost four times more likely to host a Jovian planet compared to a star with solar metallicity. Stars forming today in the disk of the Milky Way three kiloparsecs inside the solar circle have this value of [Fe/H]. From these considerations, we can infer that the incidence of giant planets must be a steep function of galactocentric distance. However, the functional form of the metallicity dependence is still poorly constrained for planet hosts more metal-poor than the Sun; the data are consistent with either a continued power-law drop-off of incidence below solar metallicity or a constant incidence in this region [98]. The lower limit on [Fe/H] for hosting a Jovian planet appears to be close to −0.5. A similar trend is not seen for Neptune-mass and smaller planets; their relative frequency appears to be independent of [Fe/H] [85,99].

More complex relations between the host star and planet properties have recently emerged from careful study of the nearby star Doppler data. However, one should be cautious in accepting these findings, as the planetary orbital parameters and even the inventory are still being updated. Dawson and Murray-Clay [100] have shown that giant planets orbiting stars more metal-poor than the Sun and having semi-major axes less than 1 AU and greater than 0.1 AU ("valley giants") tend to have more circular orbits than more metal-rich stars with giant planets. They also show that the incidence of hot Jupiters is greater around more metal-rich stars.

The incidence of Doppler-detected planets also appears to depend on the mass of the host star. Although the statistics are still weak, the current data indicate a positive correlation between planet incidence and host star mass, such that the incidence of giant planets climbs from about 2% for 0.4 $M_{sun}$ stars to about 6% for 1.4 $M_{sun}$ stars [101].

### 3.2. Exoplanets Theory

It quickly became apparent following the discovery of the first exoplanet around a sun-like star (51 Pegasi) that exoplanetary systems are generally not like the Solar System and that additional processes not previously thought to be important in the Solar System must be invoked to explain the new observed trends. To account for the hot Jupiters, like 51 Pegasi, for example, several researchers soon proposed various migration mechanisms. Migration seems to be the only way to get a Jovian planet, which presumably formed just beyond the "snow line" several AUs from the host star, to be found within 1 AU of it (Jupiter orbits 5.2 AU from the Sun). In addition, it became necessary to consider processes that result in Jovian planets having highly eccentric orbits, something unknown in the Solar System.

Planet migration remains an active, though difficult, area of research. Multiple migration mechanisms have been proposed, including gravitational interactions with the proto-planetary disk (either both the gas and planetesimal components or individually) [102,103], the Kozai mechanism, due to a distant stellar companion causing the orbit of a Jovian planet to become highly eccentric, which then becomes smaller and more circular via tidal forces from the host star [104], the Kozai mechanism, due to another planet [105], planet-planet scattering [106] and secular chaos [107].

Migration shapes the architecture of a planetary system either during the earliest stages of planet formation when gas and some solids are still present and/or later stages of planetary evolution after the disk gas has been lost. In the earliest stages of the formation of a planetary system, only gas and some solids in the form of grains are present in the disk. Then, grains begin to coalesce to form larger bodies; the bulk of the disk mass, however, is still in the form of gas. The three types of migration due to gravitational interactions between the gas disk and a planet are Type I, II and III [108]. Type I migration occurs when a planet with a small mass in relation to the disk it is imbedded within induces a spiral density wave in the disk, which, in turn, causes the planet to lose angular momentum and move closer to the star. A more massive planet has a greater effect on the disk and opens up a gap, allowing the planet and gap to migrate on the longer accretion timescale of the disk; the radial motion of the planet is coupled to the radial motion of the gas. This Type II migration is thought to be responsible for the "hot Jupiters". Type III, or runaway, migration depends on gas flow in the planet's vicinity, and movement can be very rapid either inward or outward. Migration can also occur when a planet interacts with the remnant planetesimal disk after the gas has been lost from the system.

Observational tests of particular planet migration mechanisms have involved using simulations of synthetic planetary systems that are then compared to observed exoplanet distributions (e.g., [102,109]). In other words, planet migration has to be included as part of an integrated self-consistent planetary system formation and evolution model, one which does not lend itself to a simple analytical treatment. The simulations typically include Monte Carlo methods and N-body dynamics. The models have had partial success, especially in reproducing the observed exoplanet mass and period distributions. For example, [110] correctly predicted the minimum in the planet mass distribution near 40 Earth masses for short period planets. Still, caution is urged here, as the parameters for individual systems are still being updated as new data are acquired.

One of the rare direct tests of a specific planet migration mechanism is given by [100]. They discovered that Jovian planets with orbital radii between 0.1 and 1 AU tend to have more eccentric orbits around metal-rich stars compared to metal-poor stars and interpret this as evidence that planet-planet scattering is more likely to occur in the metal-rich systems. This follows because metal-rich systems are more likely to form multiple Jovian planets that can then interact with and alter their orbits.

The mechanisms that cause planets to migrate and that increase the eccentricities of Jovian planets are related. In addition to playing a role in migration, planet-planet scattering was also proposed early on to explain the high average eccentricities of Jovian planets discovered with the Doppler method [111]. In addition, the Kozai mechanism is central to both migration and eccentricity pumping. It was recently shown that exoplanets around stars with wide binary companions tend to have higher eccentricities than exoplanets around single stars, and simulations show that the perturbing effects of the stellar companions are likely the cause of this observed difference [112]. This follows because the orbit of a distant stellar companion continuously changes, due to the changing galactic tide and impulses from nearby passing stars. Such perturbations can dramatically alter the architecture of a planetary system, even several Gyrs after its formation. This would be an interesting phenomenon to explore at different locations in the Milky Way, given the radial variation in both the strength of the galactic tides and the density of stars and interstellar clouds. It might be the case that the incidence of wide binary pairs increases with increasing distance from the galactic center.

The eccentricities of the Jovian planets in a system are relevant to the habitability of any terrestrial planets in that system. First, the variations in eccentricity of a terrestrial planet and its long-term dynamical stability depend on the eccentricities, masses and locations of the Jovian planets in a system [113]. The terrestrial planets in a planetary system containing at least one Jovian planet that is also a member of a binary star system are also influenced by the Kozai mechanism [114]. Terrestrial planets with larger eccentricities will experience greater climate changes, resulting in lower habitability [115]. Of course, other factors must also be considered when simulating the evolution of climate on a terrestrial planet, including obliquity variations, rotation period and the mass and composition of the atmosphere.

A migrating Jovian planet is likely to be a major influence on the final composition of the terrestrial planets in a planetary system. In the absence of planet migration, the condensation temperature sequence for solids in the protoplanetary disk determines the major compositional trends in a forming planetary system. It not only determines the divide between terrestrial and Jovian planets, but also the compositional differences among the terrestrial planets and among the Jovian planets. The inner planets form from refractory minerals in solids composed mainly of O, Al, Ca and Ti, while solids in the outer disk consist mainly of ices with O, Mg, Si and Fe as the major components. The formation of the Jovian planets is especially sensitive to the original metallicity of the gas (as noted above). The initial C/O and Mg/Si ratios are also important in determining the composition of the solids [116]. Jovian planet migration has the effect of redistributing solid material throughout the disk. As it migrates inward through the terrestrial planet region, material in its path either migrates along with the Jovian planet or is scattered to the outer regions of the disk. Simulations of terrestrial planet formation in the presence of Type II Jovian planet migration reveal two important results [116]. First, migration increases the fraction of terrestrial planets with a bulk composition similar to that of the Earth (mostly O, Mg, Si and Fe). Second, migration greatly increases the amount of water incorporated into the terrestrial planets, very likely resulting in water worlds.

In the absence of Jovian planet migration, the delivery of water to the terrestrial planet region from the outer disk has been found to be sensitive to the eccentricity of the Jovian planet. More eccentric Jovian planet orbits lead to a reduced water delivery to the terrestrial planets [117]. The results of these kinds of simulations are already very suggestive, but additional simulations of water delivery to terrestrial planets in systems with different architectures are required.

## 3.3. The Solar System

Consideration of these new planetary system-shaping processes has caused a revolution in our understanding of the formation and early evolution of our Solar System. With an improved understanding of the formation of the Solar System, the only known inhabited planetary system, and also the processes that form and shape exoplanetary systems, we are getting closer to understanding how a habitable planetary system forms.

It is beginning to look like some of the same mechanisms proposed to account for the observed trends among exoplanets also operated in the Solar System, albeit to a somewhat lesser degree. Since the Solar System remains the only system for which we have a complete census of the planets and a very rich census of its small bodies, it is a uniquely important source of the kind of data needed to constrain planet formation and evolution models.

The standard historical "nebular" model of the formation of the Solar System begins with the early Sun surrounded by a protoplanetary gas and dust disk, having gravitationally collapsed within a much larger interstellar molecular cloud [118]. Within this picture, the solids began condensing as small grains that coalesced to form larger solids, accumulating in the disk mid-plane. Initially, most of the mass in the disk consisted of H and He and various volatile compounds, such as water vapor and carbon monoxide. The temperature and surface mass density of the disk both decline with increasing distance from the central Sun. A very important concept is the "snow" or "frost" line, beyond which volatiles (mostly water) condensed and remained in the solid state (as ices). In the early Solar System, the snow line was between the orbits of Mars and Jupiter. Within the context of the core accretion gas-capture model [119,120], the dichotomy between the terrestrial and Jovian planets is easily explained. Jupiter formed just beyond the snow line, where the surface density of solids, mostly ices, was greatest, while the terrestrial planets formed within the snow line where solids consisted of the far less abundant refractory materials; simulations show that the terrestrial planets could have formed from a narrow annulus of material between 0.7 and 1.0 AU from the Sun [121]. Apart from a few details, such as the formation of the Moon, perhaps requiring late-time collisions, and the properties of the asteroid belt, this was considered a complete framework for understanding the formation and early evolution of the Solar System prior to the first exoplanet discovery. However, with the realization that giant planet migration must be a common phenomenon in exoplanetary systems and the application of the same general protoplanetary disk simulations to exoplanetary systems and the Solar System, it soon became apparent that even the Solar System was not immune to planetary migration. In addition, the existence of mean motion resonances among planets in some exoplanetary systems led dynamicists to reconsider their role in shaping the architecture of the Solar System.

It was in light of these developments that Tsiganis *et al.* and Gomes *et al.* [122,123] proposed the original version of the "Nice" model (Nice I), wherein Jupiter underwent inward migration and Saturn, Uranus and Neptune underwent slow outward migration, due to gravitational scattering interactions with the remnant planetesimal disk following the clearing of the gas in the protoplanetary disk. Eventually, Jupiter and Saturn passed through their mutual 2:1 mean motion resonance. This caused a dynamical instability, leading to the rapid outward scattering of Uranus and Neptune (actually exchanging their original order in the Solar System!) to their current locations, where they stabilized and circularized. Migration ceased when most of the planetesimals in the Jovian planets' zones were cleared. The motivation for proposing the Nice model included explaining the timing and magnitude of the Late Heavy Bombardment and the distributions of the asteroids and Kuiper Belt objects.

The Nice model was later revised and updated to account for other aspects of the Solar System, including additional aspects of the distribution of the main belt asteroids, the Trojan asteroids (which have the same orbit as Jupiter), the dynamical survival of the terrestrial planets and the capture of the irregular satellites by Jupiter [124–126]. This newer, Nice II, model is often called the "jumping-Jupiter" model, because the orbital period ratio between Jupiter and Saturn does not gradually pass through the 2:1 resonance, but rather jumps from less than two to greater than 2.3. The initial conditions of the Nice II model are established at the end of the gas disk phase, wherein the four Jovian planets have a compact multi-resonant configuration [127]. This gives the Nice II model a more natural initial condition than the Nice I model, which had *ad hoc* initial conditions. The capture of the Trojans and the irregular satellites both require that Jupiter had a close encounter with an ice giant [128]. This restructuring of the architecture of the outer planets resulted in Jupiter migrating inward by only a few tenths of an AU (due to its much larger mass than the other planets) and Saturn migrating outward by about 2 AU. This also opens up the possibility that the Solar System originally had a fifth ice giant, one that was lost from the Solar System when it had a close encounter with Jupiter, but the current simulations are also compatible without an extra initial outer planet.

Even with its great explanatory power, the Nice model is not able to account for some aspects of the inner Solar System. In particular, the simultaneous existence of a low mass outer terrestrial planet (Mars) and a massive terrestrial planet near 1 AU (Earth) along with an asteroid belt between two and 4 AU requires that the solid material in the early protoplanetary disk must have been truncated beyond 1 AU and then partially replenished. "The Grand tack" scenario was proposed to account for these features, envisioned as having taken place while the protoplanetary disk was still gas rich and the outer planets were still forming [129,130]. In this scenario, Jupiter would have undergone inward Type II migration to about 1.5 AU from the Sun, while Saturn was still forming, locally truncating the planetesimal portion of the disk beyond 1 AU. That Jupiter might have undergone such a large migration is surprising, but we have to remember that many Jovian planets have been observed around stars with orbital radii between one and 2 AU. Jupiter's inward migration would have continued until Saturn reached a mass near its final value and migrated inward faster than Jupiter, eventually reaching the 3:2 resonance. At this point, the two planets would have migrated outward until the disk gas was lost. The planetesimals originally in the asteroid belt region were swept clean by the first inward migration of Jupiter, but then it was replenished during Jupiter's outward migration with a much smaller population of bodies from reservoirs in the inner and outer regions of the Solar System. This naturally explains the presence of primitive volatile-rich and anhydrous parent bodies in the same narrow region of the Solar System. Given the effects of the Nice model and the Grand Tack scenario on the asteroids and other small bodies in the Solar System, it is easy to see that the asteroid belt could have been very different had the outer planets followed even modestly different histories.

In spite of the successes of the Nice model and the Grand Tack scenario (Grand Tack provides the initial conditions for Nice, so they should be considered together as one model), they suffer from some weaknesses. A number of parameters are *ad hoc*, and various free parameters are adjusted to try to match the specific properties of the Solar System rather than starting from the first principles. Examples include the mass of the initial embryos, the ratio of embryo to planetesimal mass, the accretion of the giant planets and details of the migration of the outer planets. Nevertheless, progress seems to be steady in this area.

Despite the dances of the Jovian planets, the orbits of the terrestrial planets in the Solar System have remained relatively circular. Perhaps it is this feature that most separates the Solar System from the typical exoplanet system. Long-term stability and low eccentricity of the Earth's orbit are important requirements for long-term habitability, made possible by the low eccentricities of the orbits of the Jovian planets [113]. The relative masses and positions of Jupiter and Saturn and the presence of other terrestrial planets are also relevant to the eccentricity of Earth's orbit. Had Saturn been more than about twice its actual mass, or if Saturn were closer to a major mean motion resonance or if Venus were absent, Earth's eccentricity would have been significantly larger [131,132]. Higher mass values for Saturn would cause large increases in the eccentricity of the orbit of Mars, possibly causing it to cross Earth's orbit.

The stability of Earth's obliquity is important for the maintenance of a stable climate. Laskar *et al.* [133] showed that the torque from the Moon on the Earth's equatorial bulge causes the precession frequency of its rotation axis (currently 50 arc sec/year) to be much larger than the highest secular frequency of the tilt of Earth's orbit plane (26 arc sec/year), which results from perturbations from the outer planets. Had these two frequencies been closer to being in resonance, the obliquity of Earth's rotation axis would undergo large and chaotic fluctuations; this would have been the case if Earth lacked a large Moon (but had the same rotation period) or if it rotated more slowly. Not only the amplitude, but also the rate of the obliquity variations are important for the habitability of the Earth.

How likely is it for a system like the Earth and Moon to form in the protoplanetary disk? The currently favored theory for the formation of the Moon requires a massive embryo to impact the proto-Earth embryo [134,135]; these new works show that the impact involved two bodies closer in mass to each other than had been assumed in prior work. Based on simulations with a large number of combinations of the mass of the obliquity of the planet and the mass of the satellite, Brasser *et al.* [136] find that about 2% of Earth-size terrestrial planets should form a system like our Earth-Moon system (this work also presents an informative summary of the ways the Moon likely makes Earth more habitable). Waltham [137] has noticed how close the Earth-Moon system came to being in the chaotic zone; a small decrease in the Earth-Moon angular momentum or a small increase in the Moon's mass would have caused the obliquity to be chaotic. In other words, the Moon is near its maximum mass while still avoiding chaotic obliquity variations. Waltham originally interpreted this as an anthropic selection effect on a stable obliquity and long day length. Later, Waltham [138,139] added anthropic selection for slow obliquity change as a better explanation as to why the Earth-Moon system has such a low precession frequency compared to higher, more probable values. Furthermore, the slowest obliquity variations are likely in planetary systems wherein the two most massive Jovian planets are spaced relatively farther apart, implying that the separation of Jupiter and Saturn were anthropically selected and, thus, may not be typical of Jovian planet separations around other Sun-like stars.

Water content is another very important requirement for habitability (neither too much nor too little). In addition to direct dynamical influences on the terrestrial planets, the Jovian planets also influenced the delivery of water to them. The Earth formed in a region of the early Solar System that was very dry, as evidenced by the enstatite chondrite meteorites (representative of the source bodies in the terrestrial planet region) [140]. Yet, Earth's water content is today estimated to be significantly greater than its formation at 1 AU would imply. The leading theories for the origin of Earth's water and other volatiles involve their delivery to Earth from more volatile-rich regions of the Solar System. Water delivery to Earth from comets, once a popular idea, can only account for about 10% of its crustal water inventory [140]. In the classical, pre-Grand Tack scenario, consensus delivery of volatiles from the bodies in the outer asteroid belt, perturbed by Jupiter (with a more circular orbit than present), would have been too efficient, while an orbit for Jupiter with an eccentricity comparable to the present one would have left the Earth too dry [141]. The delivery of volatiles to Earth form the outer asteroid belt region within the context of the Grand Tack scenario, however, is consistent with the measured geochemical constraints [142]. The timing of the accretion of water and other volatiles by the Earth is such that it would have occurred while it was still growing in size, but accelerating towards the final stages for its formation.

Some advances are also being made in understanding the Solar System's birth environment, though much confusion remains. The two main empirical sources of data on the very early history of the Solar System are the products of short-lived radionuclides (SLRs) in meteorites and "dynamical fossils". Observations of nearby star forming regions also give us insights into the early history of the Solar

System. Two SLRs with a long history of debate are Al-26 and Fe-60, which are believed to have been delivered to the nascent Solar System from external sources involving stellar nucleosynthesis. Various sources have been proposed, including mass loss from "super"-asymptotic giant branch stars with initial masses between seven and 11 solar masses [143], ejecta from a massive star supernova after the protoplanetary disk is formed [144], ejecta from a massive star supernova that triggers the cloud core to begin to collapse [145] and winds from Wolf–Rayet stars [146]. Several of the proposed SLR sources are estimated to be very improbable events, leading one research group to propose that the SLRs in the early Solar System came from stars formed two generations prior to it [147]. However, anthropic reasoning reminds us that a low probability for a given scenario need not disqualify it from consideration if it was a necessary step in making the Solar System habitable. A rather surprising constraint comes from recent observations of externally polluted white dwarfs, which implies that the Solar System's initial endowment of Al-26 might not have been unusual [148]. Finally, a recent reduced estimate from meteorite measurements for the initial amount of Fe-60 in the early Solar System implies that a supernova need not be invoked after all [149].

Dynamical fossils in the Solar System include the inner Oort cloud comets, Kuiper Belt objects and the unique object, Sedna; in addition, the cutoff in the Solar System planets at about 30 AU constrains the closest encounter with a star in the birth cluster. Pfalzner [150] reviewed the recent literature on this topic and considered the constraints these dynamical fossils (along with SLRs) place on the Solar System's birth cluster, finding that it most likely contained at least 1000 stars, but less than several tens of thousands of stars. Combining the SLRs constraints with their own dynamical simulations, Parker *et al.* [151] found that about 1% of the G dwarfs in their simulations are single, unperturbed and enriched in Al-26. Overall, the dynamical constraints are more informative, and future advances in this area are likely to come from the study of other objects like Sedna.

In summary, our understanding of Earth's formation has changed dramatically in recent years. The Jovian planets in the Solar System have had a complex history, and they (especially Jupiter and Saturn) have influenced the formation and evolution of the Earth and the other terrestrial planets in ways that are relevant to their habitability. With a modestly different set of initial conditions and historical trajectory, as exemplified in exoplanetary systems, the Solar System would have had markedly different Jovian and terrestrial planet architectures.

## 4. A Framework for Estimating the Number of Habitable Planets in the Universe

Astrobiology has become a very broad field, with specialists bringing along diverse knowledge and skills. At the same time, more bridges are being built, connecting previously disparate disciplines. Perhaps more so than in any other area of science, astrobiology encourages, even demands, cross-disciplinary interaction. This must happen if progress in astrobiology is going to continue. This has already been going on for a few decades in the field of Earth systems science, which seeks to understand Earth as a collection of highly interacting systems, including the biosphere, atmosphere, cryosphere, interior and nearby space environment. Long-term climate modeling for the purpose of understanding the evolution of the CHZ has been a major application of Earth systems science.

Examples of cross-disciplinary collaborations are increasing. They include climate modeling and geophysics, astrophysics and climate modeling and astrophysics and geophysics. These collaborations have been discovered to be necessary, because the historical boundaries between disciplines do not allow an astrobiology researcher to use the tools outside his area of specialty to answer some problems in astrobiology. As the region of interest widens from Earth's surface outward to the rest of the Universe, the input from astrophysics becomes ever more important, but the other disciplines will always prove to be indispensable. In other words, the cross-disciplinary collaborations, once established, must remain.

The ultimate question that most astrobiologists are seeking to answer is something like, "What is the probability that there are other planets with life?" The answer must incorporate the complete history of the Universe, including galaxy, star and planet formation and evolution. It is becoming clear that cosmology is not irrelevant to the formation and continued existence of habitable planets. The most basic elemental ingredients of planetary systems come from stars. Stars form and die in a galactic context, and galaxies form, interact and evolve in a cosmological context.

At every scale, stochastic processes shape planetary systems, and they must be modeled with Monte Carlo methods. Processes with stochastic aspects occurring on the surface of a planet include volcanic eruptions, tectonics and climate. Asteroid and comet impacts can be triggered by planetary perturbations and nearby stellar and giant molecular cloud encounters. The location and timing of specific supernovae and gamma ray bursts cannot be predicted. Encounters between galaxies can trigger star formation, threatening already-formed planets and spawning new ones.

In addition, numerical simulations are required to follow the long-term evolution of the orbits of planets around a star and the orbits of stars in a galaxy. Even with numerical simulations, however, the historical orbits of the planets in the Solar System cannot be traced back accurately more than 50 million years ago [152], and the orbit of the Sun in the galaxy cannot be traced back to its point of origin. For individual planets in a planetary system or stars in a galaxy, the orbits must be interpreted in a probabilistic sense. For these reasons, large numbers of stars and planets must be simulated and the results interpreted statistically.

The rise of interdisciplinary research in astrobiology is also revealing the complex interrelationships among the various habitability factors. For example, in the Solar System, the properties of Jupiter are relevant to the asteroid and comet impact rates on the terrestrial planets, the compositions of the terrestrial planets, the long-term orbital dynamics of the terrestrial planets and the obliquity stability of some terrestrial planets. The host star is the primary gravitational influence on the planets and affects every aspect of the dynamics of every body in a system, as well as tidal influences on the inner planets. Its electromagnetic spectrum has various effects on the atmospheres of the planets, and its particle radiation can influence atmospheric chemistry. Change one aspect of a habitable planetary system to make it non-habitable, and it might not be possible to make it habitable again with a single change to a different parameter. These aspects of a habitable environment are illustrated in Figure 3 of [1].

The considerations outlined above prevent us from estimating the probability of habitable planet formation using only analytic methods or by treating a planet in orbit around a star in isolation from the rest of the Universe. Presently, simulations of habitable planet formation are being done at two scales: the CHZ and GHZ. Several important recent studies at each scale were summarized in the present review. Possible processes that fall in the gap between the CHZ and GHZ include shrinkages of the astrosphere around a planetary system due to temporary increases in the local interstellar matter density following passages through interstellar clouds [153,154], variations in the local cosmic ray flux causing climate change on terrestrial planets [155] and the properties of a planetary system's birth environment (setting its initial conditions). These processes are better treated in a simulation that merges the CHZ and GHZ.

The next natural step is to unify the CHZ and GHZ within a single framework. The simulations would begin in the early history of a particular galaxy having particular properties. The galaxy would need to contain a large number of stars, as well as gas and dust. The chemo-dynamical evolution of the galaxy would be followed numerically, updating the star formation rate, gas and dust distributions, and the dynamics of every star at each time step. For each newly formed star, the simulation would estimate the properties of planets that form around it from its proto-planetary disk, taking into account the initial metallicity, the birth environment (e.g., loose or dense star cluster, galactic tides) and internal interactions within the system. Analytical approximations to detailed numerical simulations will need to be employed, especially in the early stages of planet formation, to keep the calculations tractable. Once the details are worked out for a spiral galaxy in isolation, then it can be placed in a broader context. For example, whether a galaxy belongs to a rich cluster or to a sparse group of galaxies will determine how often it undergoes close encounters with its neighbors. Beyond that scale, cosmological considerations can be brought to bear on the CHA and simulate the habitability of galaxies over tens of billions of years.

## 5. Conclusions

Progress in our understanding of the formation and evolution of habitable planetary systems has been remarkably rapid in recent years. Exoplanet research, in particular, is in a very healthy state, with observation and theory playing mutually supportive roles. At the same time, exoplanet research has motivated new research on the Solar System, leading to a revolution in our understanding of its formation and early evolution. Lessons learned include the following:

- It is likely that Jovian planet migration occurred in the early Solar System and influenced the formation and evolution of the terrestrial planets.
- The traditional definition of the CHZ, based on the radiant energy from the host star, is outdated and should be replaced with a definition that also includes such considerations as planetary impact rate, orbital dynamical stability and episodic reductions in the size of the astrosphere. Since these processes are partly stochastic, however, they do not lend themselves to analytic treatment.
- A planetary system cannot be isolated from its broader galactic context when considering its formation and evolution in relation to habitability. A broader and more complete understanding of habitability requires merging the CHZ and GHZ concepts.
- Even distant stellar companions can influence the dynamical stability of a planetary system several Gyrs after it formed.
- Habitability factors are often interconnected in a complex web, and some factors can have multiple distinct effects on the habitability of a planetary system. Jupiter and Saturn's influence on the habitability of the Solar System is a prime example.

## Acknowledgments

The author thanks Sarah Maddison for the invitation to write and submit this review, Andrew Howard and Erik Petigura for permission to use their figures and the anonymous reviewers for their helpful comments.

## Conflicts of Interest

The author declares no conflict of interest.